\documentclass[preprint]{aastex631}

\shorttitle{Imaging of CDG-1}
\shortauthors{van Dokkum et al.}

\begin{document}

\title{Deep HST/UVIS imaging of the candidate dark galaxy CDG-1}

\author[0000-0002-8282-9888]{Pieter van Dokkum}
\affiliation{Astronomy Department, Yale University, 52 Hillhouse Ave,
New Haven, CT 06511, USA}

\author[0000-0002-5478-3966]{Dayi David Li}
\affiliation{Department of Statistical Sciences, University of Toronto, Ontario Power Building, 700 University Avenue,
9th Floor, Toronto, ON M5G 1Z5, Canada}
\affiliation{Data Sciences Institute, University of Toronto, 700 University Avenue, Toronto, 10th floor, ON, M5G 1Z5 Canada}

\author[0000-0002-4542-921X]{Roberto Abraham}
\affiliation{David A. Dunlap Department of Astronomy and Astrophysics, University of Toronto, 50 St George Street, Toronto, ON M5S 3H4, Canada}
\affiliation{Dunlap Institute for Astronomy and Astrophysics, University of Toronto, 50 St George Street, Toronto, ON M5S 3H4, Canada}

\author[0000-0002-1841-2252]{Shany Danieli}
\affiliation{Department of Astrophysical Sciences, 4 Ivy Lane, Princeton University, Princeton, NJ 08544, USA}

\author[0000-0003-3734-8177]{Gwendolyn M.\ Eadie}
\affiliation{Department of Statistical Sciences, University of Toronto, Ontario Power Building, 700 University Avenue,
9th Floor, Toronto, ON M5G 1Z5, Canada}
\affiliation{David A. Dunlap Department of Astronomy and Astrophysics, University of Toronto, 50 St George Street, Toronto, ON M5S 3H4, Canada}
\affiliation{Data Sciences Institute, University of Toronto, 700 University Avenue, Toronto, 10th floor, ON, M5G 1Z5 Canada}

\author[0000-0001-8762-5772]{William E.\ Harris}
\affiliation{Department of Physics and Astronomy, McMaster University, Hamilton, ON L8S 4M1, Canada}

\author[0000-0003-2473-0369]{Aaron J.\ Romanowsky}
\affiliation{Department of Physics \& Astronomy, San José State University, One Washington Square, San Jose CA 95192, USA}
\affiliation{University of California Observatories, 1156 High Street, Santa Cruz CA 95064, USA}



\begin{abstract}

CDG-1 is a tight grouping of four likely globular clusters in the Perseus cluster, and a candidate dark
galaxy with little or no diffuse light. Here we provide new constraints on the luminosity of any underlying
stellar emission, using HST/UVIS F200LP imaging.
No diffuse emission is detected, with a $2\sigma$ upper limit of ${\rm F200LP}>28.1$\,mag\,arcsec$^{-2}$
on the $5\arcsec$ scale of CDG-1. This surface brightness limit
corresponds to a $2\sigma$ lower limit of $>0.5$ for the fraction
of the total luminosity that is in the form of globular clusters. The most likely alternative,
although improbable, is that CDG-1 is a chance grouping of four globular clusters in the halo of the Perseus galaxy IC 312. 
\end{abstract}




\section{Introduction} 

CDG-1\footnote{Candidate Dark Galaxy 1} ($\alpha=3^{\rm h}18^{\rm m}12.22^{\rm s}$, $\delta=41\arcdeg{}45\arcmin{}58\farcs3$)
was identified in a statistical analysis of the spatial distribution of globular clusters in the Perseus cluster.
Selecting globular clusters by their size and brightness,
\cite{li:22} found 11 overdensities in 20 HST images from the PIPER survey \citep{harris:20}. Ten 
are associated with known ultra diffuse galaxies (UDGs), but one is not: CDG-1 is comprised of 4 globular
clusters within $\approx 5\arcsec$, with no obvious smooth light component in the PIPER data \citep[see][]{li:22}.
It is therefore a candidate for a dark galaxy, or rather a galaxy whose stars are mostly or entirely locked up
in globular clusters.

The most extreme globular cluster-rich galaxies currently known are NGC\,5846-UDG1 \citep{forbes:19b,muller:21}, whose
$\approx 54$ globular clusters comprise $\sim 13$\,\% of the total light \citep{danieli:22}, and UGC\,9050-Dw1, with 
$\approx 52$ clusters making up $\sim 20$\,\% of the light \citep{fielder:23}.
As discussed
in \citet{danieli:22},
these fractions may have been close to unity at birth, because of the combined effects
of mass loss and the disruption of low mass clusters 
\citep[e.g.,][]{larsen:12,trujillogomez:19}. Turning this argument around,  CDG-1
should have some diffuse light associated with it even if all its stars originated in
globular clusters.

\section{Observations and Results}

We obtained six orbits of Hubble Space Telescope (HST) WFC3/UVIS imaging of CDG-1 to search for diffuse light
in between the globular clusters (program GO-17454). 
HST is generally not well suited for low surface brightness imaging,
but the distances between the globular clusters are so small
($1\arcsec$\,--\,$2\arcsec$) that the wings of the PSFs blend together in ground-based seeing.
The F200LP
filter was used, as this maximizes the sensitivity of HST \citep[see, e.g.,][]{kelly:22}.
The observations can be accessed via \dataset[doi:10.17909/14jb-4487]{http://dx.doi.org/10.17909/14jb-4487}.
The data were drizzled after
matching the background levels in all chips and exposures. The image is crowded, with several bright stars as
well as
the S0 galaxy IC\,312 (all with associated figure-8 ghosts). The region around CDG-1 is shown in Fig.\ 1,
with the four clusters indicated.\footnote{The numbering follows Table 3 in \citet{li:22}.}

\begin{figure*}[ht!]
\plotone{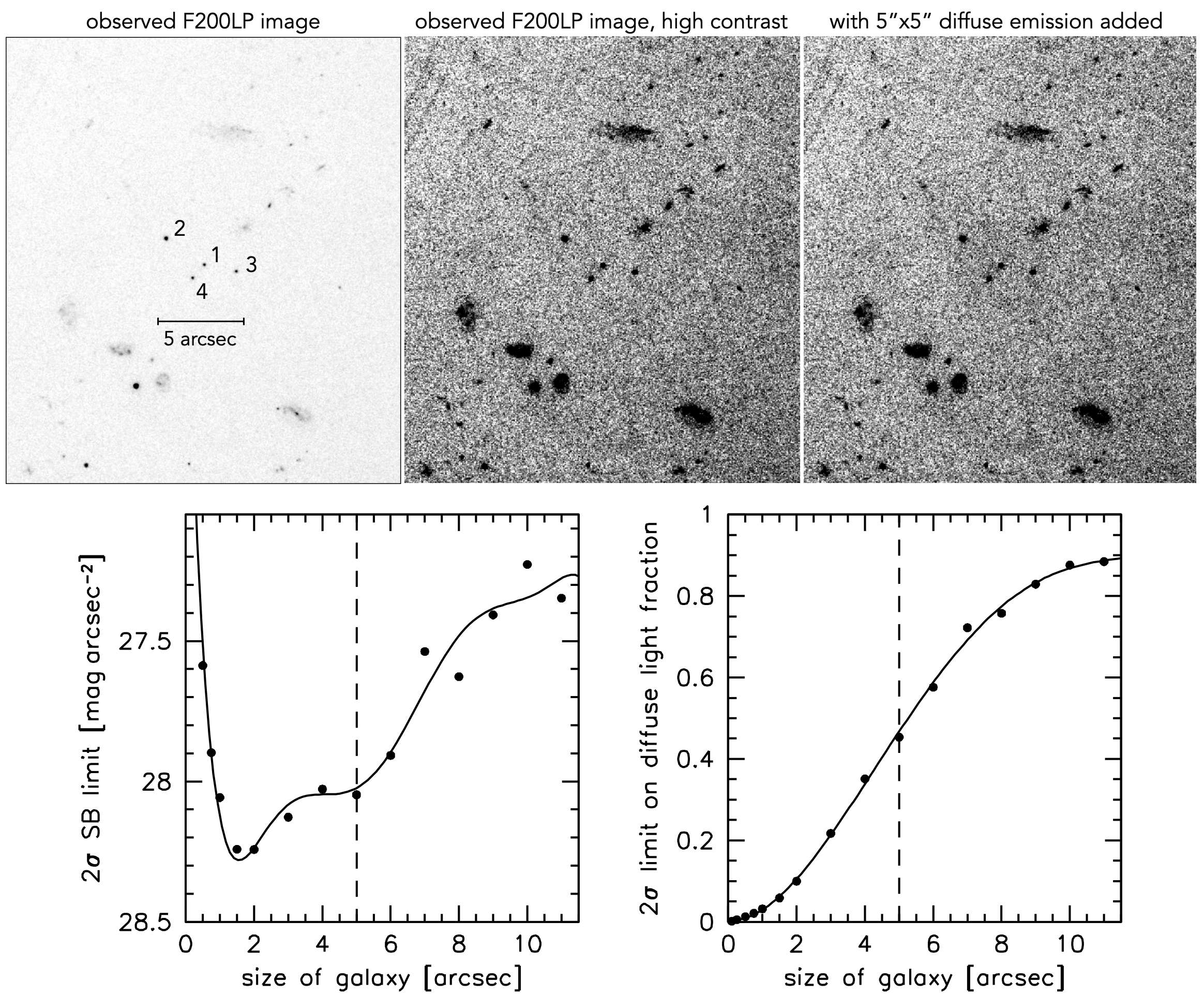}
\caption{Top left and center: section of the 6-orbit F200LP image, centered on the four globular clusters that constitute
CDG-1.
No diffuse light is detected between the clusters. Top right: same as top center, with an artificial $5\arcsec \times 5\arcsec$ patch
of diffuse light added. The patch has the same total brightness as the four globular clusters.
Bottom left: F200LP $2\sigma$ surface brightness depth as a function of spatial scale.  Bottom right:
corresponding upper limit on the fraction of the total light that is in a diffuse component.}
\end{figure*}

There is no significant detection of diffuse light between the globular clusters; neither median filtering nor binning nor Gaussian
smoothing shows a flux enhancement with respect to the surroundings. The scale-dependent empirical
$2\sigma$ surface brightness limit, as determined from an object-masked image
with the {\tt sbcontrast} code \citep{keim:22}, is shown in the bottom left panel of Fig.\ 1 (after
applying a 0.5\,mag Galactic extinction correction). On the $\approx 5\arcsec$ scale
of CDG-1 the F200LP limit is $28.1$\,mag\,arcsec$^{-2}$ (AB). 
Using the measured F200LP fluxes of the globular clusters and multiplying the surface brightness by the square of
the spatial scale to get total diffuse magnitudes, we can calculate the maximum fraction of the total
light that is in a diffuse component (bottom right panel in Fig.\ 1). We find a $2\sigma$ upper limit of $\approx 0.5$ for
this fraction for a galaxy size of $5\arcsec$ (1.8\,kpc at the 75\,Mpc distance of Perseus). The top right
panel in Fig.\ 1 illustrates what an artifical diffuse object
of this size and brightness ($m=24.6$) looks like; as expected, it is just visible.
The limit on the fraction is a strong function
of size: for larger sizes, the per-arcsec$^2$ surface brightness limit is weaker, and the total brightness of the diffuse component is
larger by a factor (size)$^2$.

\section{Discussion}

The apparent lack of diffuse emission in between the globular clusters
is, at face value, difficult to reconcile with the notion that CDG-1 is a galaxy.
It requires that no stars were formed outside of the four globular clusters and that there was no appreciable mass loss
or destruction of low mass clusters \citep[see][]{danieli:22}. It is possible, however, that the diffuse emission has a larger spatial
extent than the region occupied by the four globular clusters. Specifically, if the galaxy
spans $10\arcsec\times 10\arcsec$
(3.6\,kpc\,$\times$\,3.6\,kpc, a modest-sized UDG),
the UVIS data allow diffuse light contributions up to $\approx 90$\,\% of the total light (see Fig.\ 1).

An alternative explanation is that CDG-1 is not a galaxy. It is unlikely that the grouping occurred by chance in
the general Perseus cluster field \citep{li:22}. However, the four clusters might be part of the
outer globular cluster population of IC\,312 \citep[see
the discussion in][]{li:22}.  In conclusion, the UVIS data do not definitively tell us whether or not CDG-1 is a dark galaxy.
Radial velocities, and the possible detection of other examples \citep{li:23}, may provide further constraints.

\bibliography{master}{}
\bibliographystyle{aasjournal}

\end{document}